\documentclass[prl,twocolumn,aps,superscriptaddress]{revtex4-1}
\usepackage{graphicx}
\usepackage{amsmath}
\usepackage[colorlinks=true,citecolor=blue, urlcolor=blue, linkcolor = blue]{hyperref}
\usepackage{amssymb}
\usepackage{cancel}
\usepackage[dvipsnames]{xcolor}
\usepackage{float} 
\usepackage[caption=false]{subfig}
\definecolor{lgray}{gray}{0.8}
\usepackage{verbatim} 
\usepackage{tabularx} 
\newcolumntype{Y}{>{\centering\arraybackslash}X} 
\usepackage{braket}

\begin{document}

\title{Spatiotemporal dynamics of particle collisions in quantum spin chains}

\author{P. I. Karpov}
\email{karpov@pks.mpg.de}
\affiliation{Max Planck Institute for the Physics of Complex Systems, N{\"o}thnitzer Stra{\ss}e 38, Dresden 01187, Germany}
\affiliation{National University of Science and Technology ``MISiS'', Moscow, Russia}

\author{G.-Y. Zhu}
\email{guoyi@pks.mpg.de}
\affiliation{Max Planck Institute for the Physics of Complex Systems, N{\"o}thnitzer Stra{\ss}e 38, Dresden 01187, Germany}

\author{M. P. Heller}
\email{michal.p.heller@aei.mpg.de}
\affiliation{Max Planck Institute for Gravitational Physics, Am M{\"u}hlenberg 1, 14476 Potsdam, Germany}
\altaffiliation{\emph{On leave from:} National Centre for Nuclear Research, Pasteura 7, Warsaw, 02093, Poland}
\altaffiliation{}

\author{M. Heyl}
\email{heyl@pks.mpg.de}
\affiliation{Max Planck Institute for the Physics of Complex Systems, N{\"o}thnitzer Stra{\ss}e 38, Dresden 01187, Germany}
\date{\today}
\begin{abstract}
Recent developments have highlighted the potential of quantum spin models to realize the phenomenology of confinement leading to the formation of bound states such as mesons.
In this work we show that Ising chains also provide a platform to realize and probe particle collisions in pristine form with the key advantage that one can not only monitor the asymptotic particle production, but also the whole spatiotemporal dynamics of the collision event.
We study both elastic and inelastic collisions between different kinds of mesons and also more complex bound states of mesons, which one can interpret as an analog of exotic particles such as the tetraquark in quantum chromodynamics.
We argue that our results not only apply to the specific studied spin model, but can be readily extended to lattice gauge theories in a more general context.
As the considered Ising chains admit a natural realization in various quantum simulator platforms, it is a key implication of this work that particle collisions therefore become amenable within current experimental scope.
Concretely, we discuss a potentially feasible implementation in systems of Rydberg atoms.
\end{abstract}

\maketitle

\noindent \emph{Introduction.--} Particle collisions of hadronic matter, such as of protons or nuclei at the LHC~\cite{Campana:2016cqm,Busza:2018rrf}, represent a key element to probe fundamental forces.
Hadronic matter forms due to the confinement of quarks leading to the formation of bound states, as described by a (1+3)-dimensional SU(3) gauge theory -- quantum chromodynamics (QCD)~\cite{Shuryak:2004pry}.

The quest of solving QCD has triggered many promising recent developments at the interface of high-energy and quantum many-body physics. What directly motivates our efforts ultimately descends from the pioneering work~\cite{McCoy:1978} and concerns studying the effects of confinement in quantum spin chains~\cite{Greiter:2002,Lake:2010,Morris:2014,Grenier:2015, Kormos:2017,Bera:2017, Lerose:2019jrs, Mazza:2019, Robinson:2019, James:2019,Banuls:2019qrq, Vanderstraeten:2020,Lerose:2020,Verdel:2020}. Related advances concern studying lattice gauge theories using quantum information techniques, as reviewed in Refs.~\cite{Banuls:2019bmf,Banuls:2019rao}.

Triggered by these developments, as well as by the importance of collisions in high-energy physics, our work initiates studies of particle collisions in paradigmatic quantum matter.
There are two key advantages of our setup. First, it allows us to monitor not only the asymptotic particle production, but also the whole spatiotemporal dynamics of the full collision event in pristine form. Second, our setup lies within reach of quantum simulators, which have already explored the influence of confinement onto quantum many-body dynamics~\cite{vovrosh2020confinement,Tan:2019kya}.

\begin{figure}[tbh] 
	\centering
	\includegraphics{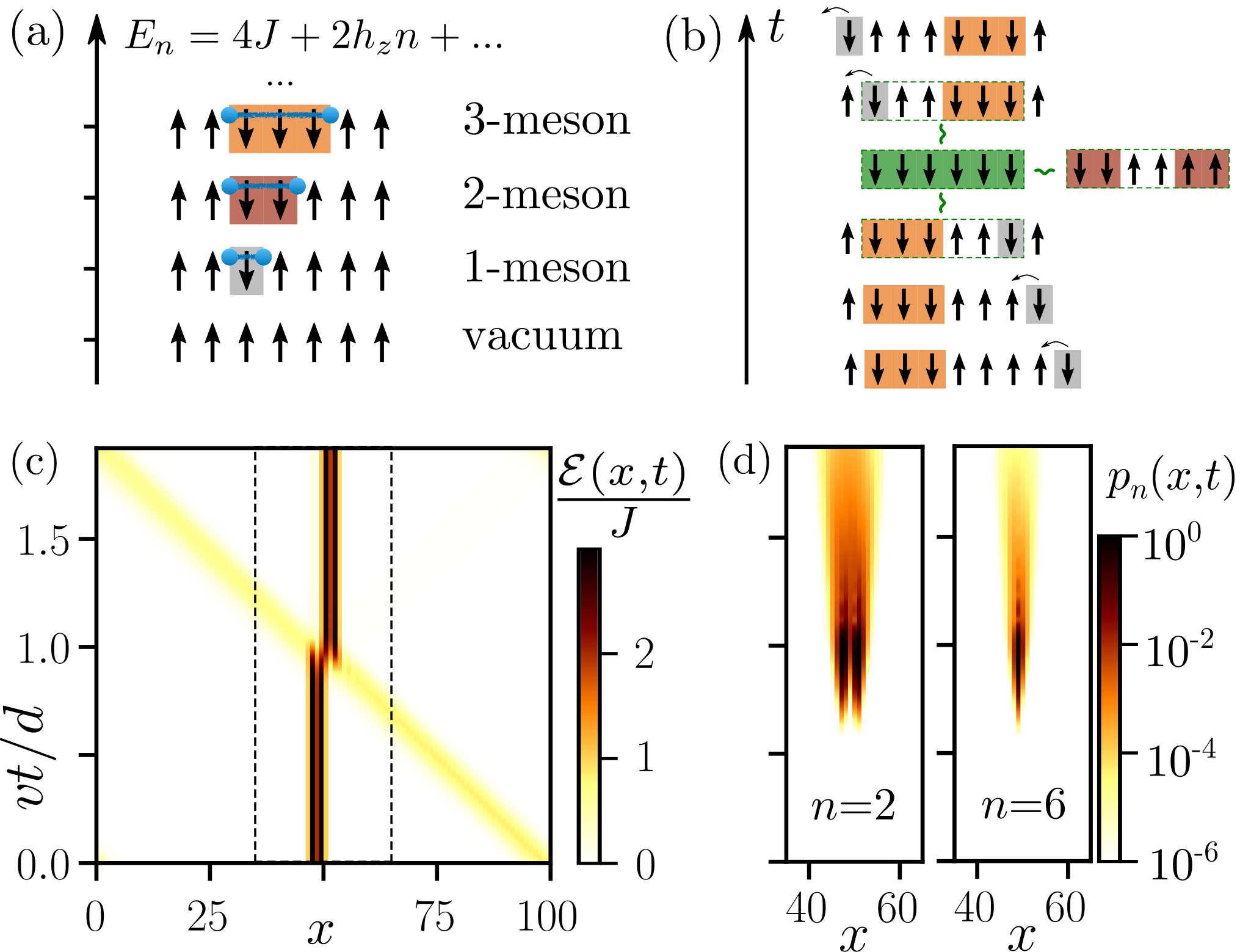}
	\caption{
	    (a) Low energy states of the Ising chain:  domain walls (blue dots) are confined in pairs (``$n$-mesons'') due to a linear interaction potential $E_n\sim h_z n$, where $n$ denotes the separation between the domain walls.
	    (b) Schematic illustration of the $3+1$-collision of a $3$-meson with a $1$-meson for $h_z=J$ highlighting the formation of intermediate $6$- and $2$-mesons.
	    (c,d) Spatiotemporal picture of the $3+1$-collision for an initial distance $d$ and a $1$-meson wave packet with momentum $k=\pi/2$ and velocity $v=2h_x^2/(3J)$.
	    (c) Local energy $\mathcal{E}(x,t)$. (d) Spatially-resolved meson occupations $p_n(x,t)$~\eqref{eq.spatialp} for $n=2$ (left) and $n=6$ (right). The data is obtained for a chain of $L=100$ sites within the $2^\text{nd}$ order effective model.
	}
	\label{fig:3+1_spatiotemporal}
\end{figure}

In more concrete terms, we explore here the dynamics generated by quantum Ising chains, where confinement of domain walls can be induced by longitudinal fields~\cite{Kormos:2017}, for initial conditions which realize mesonic wave packets impacting onto each other, see Fig.~\ref{fig:3+1_spatiotemporal}.
We analyze the influence of projectiles on the outcome of collisions and discover, in particular, a possibility of particle production during inelastic collisions. We also discuss a potential implementation and its feasibility in the context of Rydberg atoms~\cite{Bernien:2017,Zeiher:2017,Marcuzzi:2017,Lienhard:2018,Guardado:2018,Leseleuc:2018}.

\noindent \emph{Setup.--}
We consider the quantum Ising chain with both transverse~$h_{x}$ and longitudinal~$h_{z}$ fields
\begin{align}
H = -J \sum_{i=1}^L \sigma_i^z \sigma_{i+1}^z - h_x \sum_{i=1}^L  \sigma_i^x  - h_z \sum_{i=1}^L  \sigma_i^z,
\label{eq:hamiltonian}
\end{align}
where $\sigma_i^{x(y)(z)}$ denotes the Pauli matrix at site $i$ in a periodic chain of $L$ sites.
The presence of the longitudinal field splits the otherwise degenerate Ising vacuum in the ferromagnetic phase with the central consequence that two domain walls now experience an interaction potential increasing linearly as a function of their distance~\cite{Kormos:2017}.
This leads to confinement and imposes a string tension between domain walls.
The low-lying excitations are bound states of two domain walls, connected by an electric string with energy proportional to the string length, see Fig.~\ref{fig:3+1_spatiotemporal}(a). The Ising model can be mapped to a $\mathbb{Z}_2$ lattice gauge theory with $\sigma_j^z$ as the electric field and additional $\mathbb{Z}_2$ matter charges at each domain wall~\cite{Lerose:2019jrs}. Therefore, not only are the bound states rightfully called mesons, as $\mathbb{Z}_2$ analogues of SU(3) ones in QCD, but also our analysis applies to the broader context of lattice gauge theories.

For the purpose of obtaining full spatiotemporal resolution of collisions we focus on the limit of a weak transverse field, i.e. $h_x\ll J$.
We derive an effective description in terms of the aforementioned mesons by defining hard-core bosonic operators $\psi_x^{[n]\dagger} =P_{x+[n/2]-n}^\uparrow \prod_{j=x+[n/2]-n+1}^{x+[n/2]}\sigma_j^-  P_{x+[n/2]+1}^\uparrow$ with  $P_{j}^{\uparrow(\downarrow)}\equiv (1\pm \sigma_j^z)/2$ the projector onto $\uparrow(\downarrow)$ at site $j$. 
By means of a Schrieffer-Wolff transformation (SW)~\cite{Bravyi_2011} applied to Eq.~(\ref{eq:hamiltonian}) for weak $h_x$ we arrive at:
\begin{equation}
\label{eq.Heff}
    H_{\text{eff}} = H_0 + H_{\text{int}}  \, ,
\end{equation}
where $H_{0}$ for the case of the most interest to us, $h_z\simeq J$, reads
\begin{equation}
\label{eq.H0}
   H_0 = \sum_{n,x} m_n p_n(x) - \frac{v}{2}\sum_x \left(\psi_x^{[1]\dagger} \psi_{x+1}^{[1]}+h.c.\right)
\end{equation}
and contains both the energies of the $n$-mesons,
\begin{equation}
\label{eq.energiesmesons}
m_n=4J+2n h_z +(4-n-4\,\delta_{n,1})h_x^2/3J+O(h_x^4),
\end{equation}
and the velocity of the $1$-mesons
\begin{equation}
\label{eq.vdef}
v=\frac{2h_x^2}{3J}.
\end{equation}
Here, 
\begin{equation}
\label{eq.spatialp}
    p_n(x)=\psi_x^{[n]\dagger}\psi_x^{[n]}
\end{equation}
denotes the occupation of a $n$-meson at site $x$. 
Other values of $h_z$ would modify the masses and velocities.
In the regime we are interested in, $h_x\ll h_z\simeq J$, only $1$-, $2$- and $3$-mesons are quasi-stable particles. 
The Hamiltonian in Eq.~(\ref{eq.H0}) gives a dispersion relation for the lightest meson according to $\epsilon_k=m_1-v\cos k+O(h_x^4)$.
The heavier $n$-mesons with $n>1$ acquire a kinetic term only at higher orders in perturbation theory which can be neglected on the time scales $t\ll h_x^4/J^3$ we consider. As a result, $m_{n}$'s with $n>1$ act as rest masses, whereas $m_{1}$ is the maximal kinetic energy of the $1$-meson.

Essential for the targeted particle collisions, the mesons also exhibit interactions,
\begin{equation}
\begin{aligned}
\label{eq.Hint}
&H_{\text{int}} = -\frac{h_x^2}{J} \sum_{m,n,x} \left(\psi_x^{[m+n+2]\dagger} \psi_{x}^{[m]}\psi_{x+m+2}^{[n]}+h.c.\right)\\
&+\frac{h_x^2}{J}\sum_{m,n,x} 
\left(\psi_{x}^{[m-1]\dagger}\psi_{x+m}^{[n+1]\dagger}
\psi_{x}^{[m]}\psi_{x+m+1}^{[n]}+h.c.\right)\\
&+\frac{3h_x^2}{2J}\sum_{m,n,x} 
\left(\psi_{x}^{[m]\dagger}\psi_{x}^{[m]}
\psi_{x+m+1}^{[n]\dagger}\psi_{x+m+1}^{[n]}+h.c.\right),
\end{aligned}
\end{equation}
which we again displayed for $h_z=J$. The first interaction term describes the fusion of two close-by mesons $m$ and $n$ to a heavier $m+n+2$ one upon converting domain wall excitations into string energy, or the reverse process (essential for string breaking). The second term  describes string exchange between two nearby mesons. The last term is a repulsive nearest-neighbor density-density interaction between two mesons.

Upon considering a general $h_z\not= J$ only slight modifications have to be incorporated in Eq.~(\ref{eq.Hint}) such as $h_z$-dependent corrections to the coupling constants~\cite{supplement}. Also the fusion interaction would have to be skipped~\cite{supplement}.

\begin{figure}[t] 
	\centering
	\includegraphics[width=1.0\linewidth]{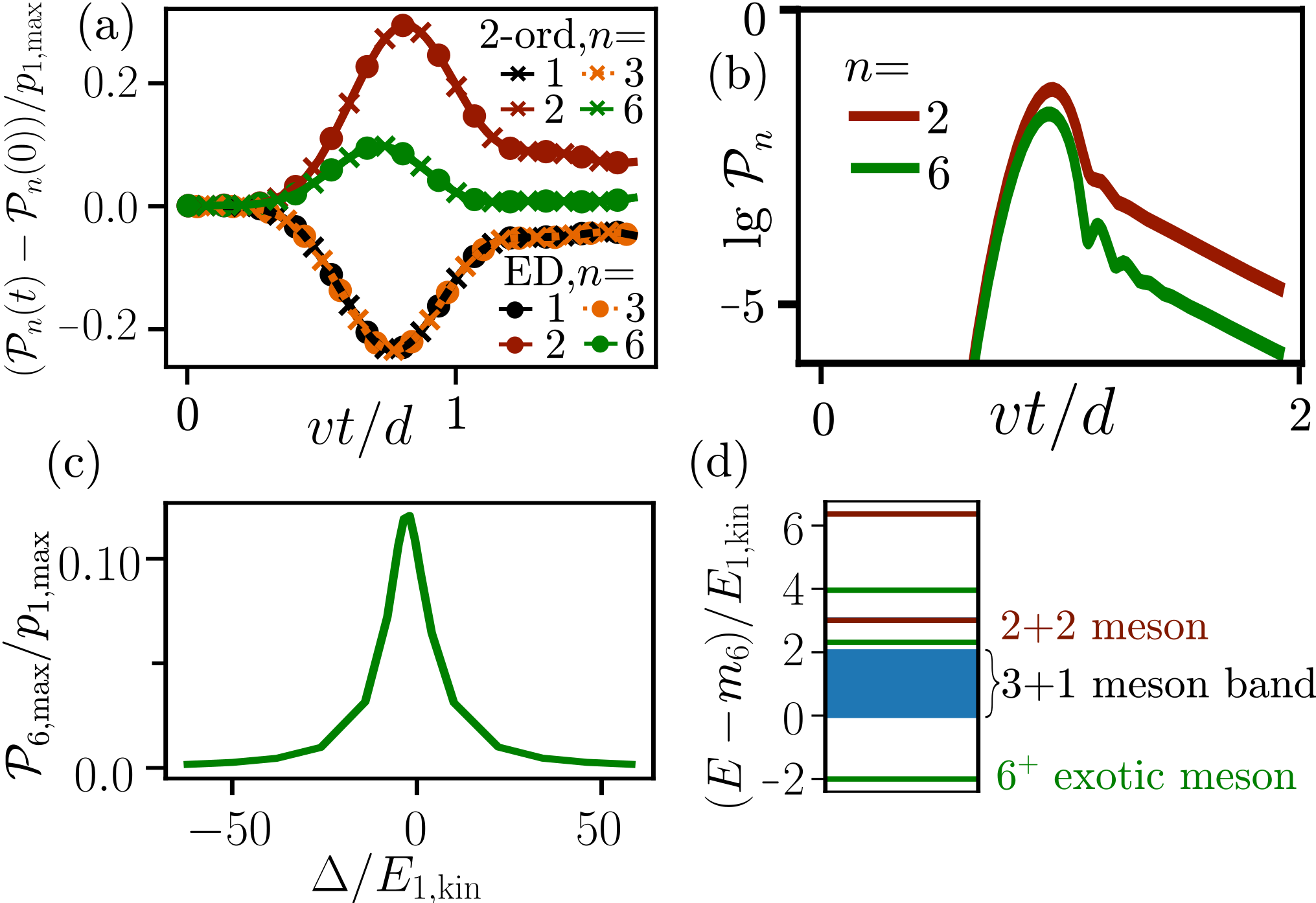}
	\caption{
	Particle production in the $3+1$-collision for $h_z=J$ and $h_x=10^{-2}J$. (a) Normalized particle production, $(\mathcal{P}_n(t)-\mathcal{P}_n(0))/p_{1,\mathrm{max}}$, comparing ED of the full Ising model to the effective theory for $L=20$; $p_{1,\mathrm{max}}$ is the maximum on-site probability of the 1-meson wave packet, $v=2h_x^2/(3J)$ its group velocity, and $d=L/2$ the initial distance between the centers of 3- and 1-mesons.
		(b) Dynamics of the $2$- and $6$-meson occupations for $L=100$ calculated for the effective model.
		(c) Maximum production of the intermediate states vs detuning $\Delta$~\eqref{eq.detuning} normalized by the kinetic energy $E_{1,\mathrm{kin}}$ of the $1$-meson.
		(d) Spectrum in the vicinity of the exotic $6^+$-meson.
	}
	\label{fig:3+1_projectors}
\end{figure}

The limit $h_x \ll J$ has the particular advantage that the mesons become spatially localized, which allows us to access the full spatiotemporal resolution of the collision dynamics.
In the SW rotated basis all the $n$-meson occupations can be directly measured from simple projective measurements in the $\sigma^z$ basis.
In the original unrotated basis the meson operators are dressed upon inverting the SW.
Importantly, this dressing is only of perturbative nature in $h_x/J$ as opposed to the collision event itself so that in the limit $h_x/J \to 0$  upon keeping $h_x^2t/J=\text{const.}$ the meson expectation values as computed with $H$ and $H_\text{eff}$ become identical, see Fig.~\ref{fig:3+1_projectors}(a).

We complement the studied meson observables by monitoring further the local energy ${\cal E}(x,t)$ at site $x$ being the SW transformed Ising Hamiltonian density $-J \sigma_i^z (\sigma_{i-1}^z+\sigma_{i+1}^z)/2 - h_x  \sigma_i^x  - h_z \sigma_i^z$ with $i \equiv x$.

We study collisions by employing exact diagonalization (ED) for both the full model in Eq.~(\ref{eq:hamiltonian}) for moderate $L$ and the effective Hamiltonian in Eq.~(\ref{eq.Heff}) for large $L$ with the former corroborating the asymptotic exactness for the latter calculations for $h_x/J \to 0$.

\noindent \emph{Collision protocol.--}
Based on the knowledge of the mesonic excitations, it is straightforward to realize a collision scenario.
We can generate propagating $n$-mesons in the form of Gaussian wave-packets by the operator
\small
\begin{equation}
\label{eq:psin}
\psi^{[n]}(x_0,k_0)^\dagger=\frac{1}{\sqrt{\mathcal{N}}}\sum_{x=-\infty}^{\infty} e^{-x^2/(4\tau_x^2)} e^{i k_0 x } \prod_{j=0}^{n-1}\sigma_{x_0+x+j}^-,
\end{equation}
\normalsize
acting on the ferromagnetic background,with $\tau_x$ denoting the width and $\mathcal{N}$ the normalization factor.
The characteristics of the collision process, however, do not depend on the details as long as  the wave packet is sufficiently localized in momentum space so as to avoid any spreading on the considered time scales.
Concretely, we choose $\tau_x=\sqrt{L/4\pi}$ implying $\tau_k \sim 1/\sqrt{L}$ for the width in momentum space.
Practically, for the collision we decompose the system into two halves each of which contains one of the two colliding particles and we constrain the summation over $x$ to just one half so as to avoid any overlap of the two initial mesons.
In the limit $h_x\ll J$, only the $1$-meson can propagate on time scales $t\ll J^3/h_x^4$, see Eq.~(\ref{eq.H0}), so that we focus on collisions involving at least one $1$-meson, while also collisions with propagating higher-meson states can be realized when addressing longer time scales~\cite{supplement}.
We maximize the kinetic energy by considering $1$-mesons at maximum group velocity by choosing $k_0 = \pm\pi/2$.
Static higher $n$-mesons can be prepared by just flipping $n$ spins in an otherwise ferromagnetic background.

As we will discuss in detail in the following, collisions of a variety of different characters including elastic and inelastic ones with both heavy mesons and exotic mesonic bound states can be realized in the considered model.
In particular, in inelastic collisions the kinetic energy of the incident particles is converted into the creation of new (with respect to colliding particles) meson states.

We find that the mesonic spectrum provides key information about the nature of the collision process, which we use to controllably tune and study different types of collisions.
At an analytic level the effective Hamiltonian in Eq.~(\ref{eq.Heff}) can be used as an insightful starting point.
Focusing on the rest masses $m_n$, one can directly see that it is straightforward to realize elastic collisions where the only energy conserving process is elastic scattering of the incident particles without a resonant channel to generate final meson states different from the initial ones.
When it comes to more complex inelastic collisions we find that the full spectra of the Hamiltonians in Eqs.~(\ref{eq:hamiltonian},\ref{eq.Heff}) are very valuable for their understanding, but also to predict interesting events.

\noindent \emph{Inelastic collision.--}
For the purpose of studying inelastic collisions we focus on a strong coupling limit $h_z=J\gg h_x$ implying a large string tension, which turns out to be particularly fruitful for the purpose of identifying resonant channels.
For this parameter regime, a state consisting of a $3$- and a $1$-meson is resonant with both a $6$-meson and a state involving two $2$-mesons from an analysis of the rest masses in the effective model of Eq.~(\ref{eq.Heff}) upon neglecting for a moment the perturbative $(h_x/J)^2$ corrections.

In Fig.~\ref{fig:3+1_spatiotemporal}(b) we show a schematic picture of the collision process with a $1$-meson incident on the $3$-meson.
When the target and the incoming meson are separated by only two $\uparrow$-spins, a second-order in $h_x/J$ spin flip process, i.e., the fusion interaction term in Eq.~\eqref{eq.Hint}, maps them to a resonant (based on the estimate of the classical rest masses without $h_x/J$ corrections) $6$-meson configuration.
The $6$-meson can then transform resonantly via a similar process of flipping the two central spins to a state with two $2$-mesons or back to a configuration with a $1$-meson on the left of a $3$-meson.
These considerations based on the leading order expression for the rest masses $m_n$ already provide the basic picture for the actual collision dynamics during the full nonequilibrium quantum evolution.
There is, however, also an alternative channel via the string exchange interaction term in Eq.~(\ref{eq.Hint}) that is present irrespective of the value of $h_z/J$: whenever a $1$-meson is the nearest neighbour of the $3$-meson, the single $\uparrow$ spin between them can hop to the left through a resonant two-spin flip resulting in two nearest neighbouring $2$-mesons: $\ket{\downarrow\downarrow\downarrow\uparrow\downarrow}\rightarrow\ket{\downarrow\downarrow\uparrow\downarrow\downarrow}$. When the single $\uparrow$ spin further moves to the left by a further resonant exchange process, the $1$- and $3$-meson are restored with the exchanged positions. More details on the full set of resonant process in this collision can be found in~\cite{supplement}.

The spatiotemporal dynamics of the local energy $\mathcal{E}(x,t)$ in Fig.~\ref{fig:3+1_spatiotemporal}(c) allows one to clearly identify the shift of the $3$-meson during the collision, which one can already see from the schematic picture in Fig.~\ref{fig:3+1_spatiotemporal}(b) and which might be viewed as a particle collision analog of the Newton’s cradle.
Figure~\ref{fig:3+1_spatiotemporal}(d) shows the space-time particle production of the $6$- and $2$-mesons during the collision process.
After the collision, the $6$-meson and a pair of $2$-mesons can effectively propagate mediated by the motion of $1$-meson, while gradually decaying back to $3+1$ meson states~\cite{supplement}.

In Fig.~\ref{fig:3+1_projectors} we analyse the individual $n$-mesons across the collision based on their global occupations \begin{equation}
    \mathcal{P}_n(t) = \sum_x p_n(x,t).
\end{equation}
In Fig.~\ref{fig:3+1_projectors}(a) we display $\mathcal{P}_n(t)$ relative to the initial condition $\mathcal{P}_n(t=0)$, in order to highlight the changes in occupation due to the collision.
We further properly normalize the particle production to the height $p_1^\text{max} = \max_x p_1(x,t=0)$ of the incident $1$-meson wave packet.
With this we eliminate a dependence on the details of the incident wave packet structure making the particle production directly comparable between differently initialized wave packets.
A wave packet broader in real space will hit the $3$-meson only with a reduced maximal amplitude thereby also lowering the particle production at a given instant in time.
In Fig.~\ref{fig:3+1_projectors}(a) we include both data from full exact diagonalization (ED) for Eq.~(\ref{eq:hamiltonian}) as well as from the effective Hamiltonian in Eq.~(\ref{eq.Heff}) for $h_x/J=0.01$.
We achieve a collapse of the two data sets confirming the accuracy of Eq.~(\ref{eq.Heff}), which we will therefore use extensively in the remainder of this work.

Figure~\ref{fig:3+1_projectors}(a) shows that the collision is accompanied by a significant production of $2$- and $6$-mesons at the expense of a reduction of the $1$- and $3$-mesons accordingly, which follows the considerations along the lines of the schematic picture in Fig.~\ref{fig:3+1_spatiotemporal}(b).
After the collision some of the intermediate high $2$- and $6$-meson occupations convert back to $1$- and $3$-mesons.
Importantly, also an inelastic channel remains as we display in more detail in Fig.~\ref{fig:3+1_projectors}(b) for $2$- and $6$-meson occupations on a logarithmic scale.
As one can see, these particles are not just created via some intermediate state but remain also after the collision as the inelastic contributions with a weak decay over time.
Note that the time axis in Fig.~\ref{fig:3+1_projectors}(b) is rescaled by the initial distance $d$ implying a long lifetime in the bare microscopic units.
The decay of such high-energy particles is in line with general expectations of string breaking suggesting that heavy mesons with a large string tension such as the $6$-meson decay on the expense of creating new lighter particles~\cite{Lerose:2020,Verdel:2020}.
Up to this point we have chosen $h_z=J$ to achieve a resonance based on an estimate of the bare rest masses leading to the natural question of what happens upon tuning the system out of this resonance.
In Fig.~\ref{fig:3+1_projectors}(c) we show the dependence of the particle production of the $6$-meson as a function of the detuning of the energy barrier
\begin{equation}
\label{eq.detuning}
\Delta = m_6 - m_1 - m_3 = 4(h_z-J)-\frac{2}{3}\frac{h_x^2}{J}.
\end{equation}
One can identify a clear Lorentzian-type resonance peak, as one might expect from conventional collision processes.
The particle production for the $2$-meson includes two channels during the collision, which compete with each other yielding a subtle interplay.
While one channel shows a resonance analogous to the $6$-meson at the same~$\Delta$, the other resembles an anti-resonance due to this competition, for a more detailed discussion see~\cite{supplement}.

In order to understand, why the inelastic channel is not more effective in creating a larger production of, e.g., the $6$-mesons, it is very instructive to go one step further by studying the full many-body spectra instead of just the rest masses as done before.
In Fig.~\ref{fig:3+1_projectors}(d) we show the energy levels of the Hamiltonian in Eq.~(\ref{eq.Heff}) diagonalized in the zero-momentum sector normalized to the kinetic energy $E_{1,\text{kin}}=2h_x^2/(3J)$ of the incident $1$-meson.
As one can see, interactions induce significant modifications.
Most notably, the $6$-meson evolves into a set of three exotic meson states $6^+$ indicated by the green lines in Fig.~\ref{fig:3+1_projectors}(d).
Specifically, the $6^+$ states are bound states of various elementary $n$-mesons~\cite{supplement}, which one might interpret as an analog of exotic mesons such as the tetraquark in QCD~\cite{Liu:2019zoy}.
We will make use of the short-hand notation $6^+=6+1\otimes 3 + 2\otimes 2$ to denote a superposition state composed out of a $6$-meson, a product states of a $1$- and a $3$ meson, as well as a product state of two $2$-mesons, all of which with some amplitudes, whose exact values are not of concrete importance but can found in~\cite{supplement}.
Most importantly, the $6^+$ exotic mesons are pushed out of the continuum made up of $1+3$-mesons, which contains the particle configurations before the actual collision event.
Therefore, the transition to the $6$- and $2+2$-meson states becomes slightly off-resonant, which provides an explanation of why their respective production only reaches a value of $\mathcal{P}_n/p_1^\text{max} \approx 0.1$, see Fig.~\ref{fig:3+1_projectors}(a).

\begin{figure}[tb] 
	\centering
	\includegraphics{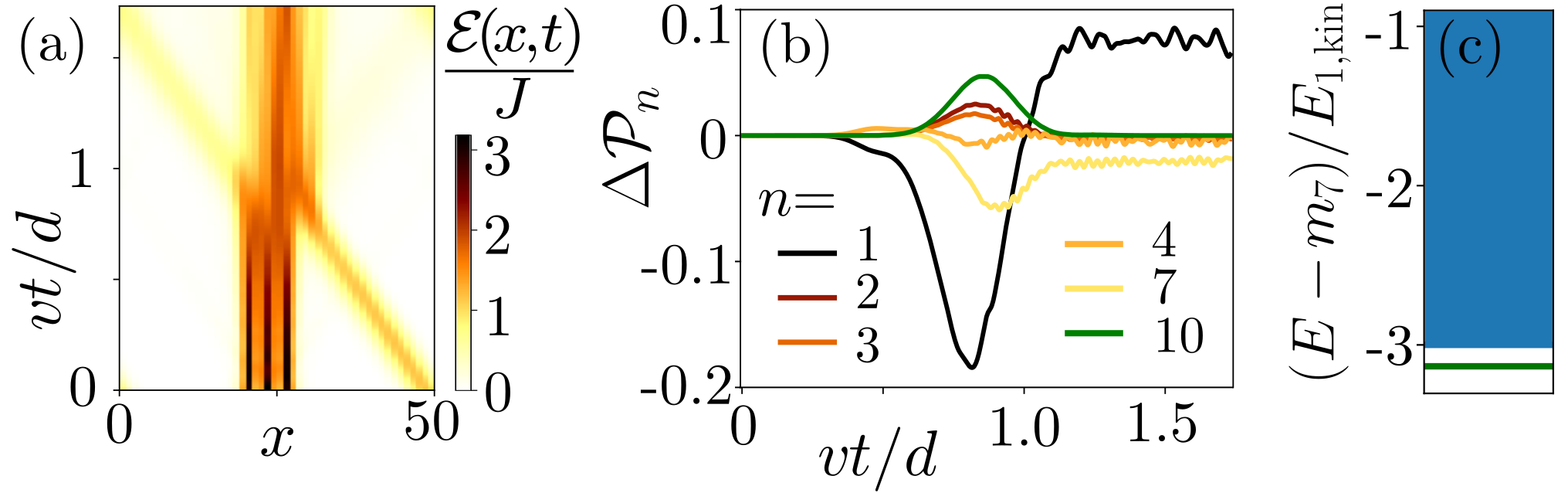}
	\caption{
	Inelastic collision between the exotic $7^+$-meson state and a $1$-meson.
		(a) Spatiotemporal profile of the local energy $\mathcal{E}(x,t)$.
		(b) Change of the meson occupation $\Delta\mathcal{P}_n(t) = \mathcal{P}^{\mathrm{coll}}_n(t)-\mathcal{P}^{7^+}_n(t) - \mathcal{P}^{1}_n(t)$ due to the collision; here $\mathcal{P}^{7^+}$ and $\mathcal{P}^{1}$ correspond to single $7^+$- and 1-meson respectively, while ${P}^{\mathrm{coll}}$ corresponds to $7^+$ and $1$ collision.
		(c) Zoom into the many-body spectrum close to the exotic $7^+$-meson in the zero-momentum sector (green), which is separated from the three $1$-meson continuum (blue) by a small energy gap $\sim 0.1h_x^2$.
     	}
	\label{fig:bound_state}
\end{figure}

\noindent \emph{Inelastic collision with exotic meson.--}
Having established inelastic collisions in the Ising model between $1$- and $3$-mesons we now aim to take one further step towards collisions with more complex objects.
In this context it is of particular importance that from the spectra, see Fig.~\ref{fig:bound_state}(c), we can also identify an exotic meson bound state $7^+=7+1\otimes 1\otimes 1$ of a $7$-meson with three $1$-mesons, which  is located slightly below the continuum consisting of three $1$-mesons.
Crucially, however, the gap is small compared to the kinetic energy of an incident $1$-meson leaving us with the expectation that a collision between the exotic $7^+$-meson state and a $1$-meson could excite the $7^+$ into the continuum $1$-meson band.
In Fig.~\ref{fig:bound_state}(a) we show the spatiotemporal dynamics of the local energy $\mathcal{E}(x,t)$ for a setup with a $1$-meson projectile impacting onto a $7^+$ state.
The initial preparation of the $7^+$ state we achieve based on the knowledge of the exact solution by creating a local superposition state of a $7$ meson and three $1$-mesons with amplitudes set according to the respective eigenstate, for the numerical details of the amplitudes see~\cite{supplement}.
Compared to the previous inelastic collision, see Fig.~\ref{fig:3+1_spatiotemporal}, the present one appears more violent changing significantly the real-space structure.
The impact of the $1$- onto the exotic meson creates a complex transient state involving the production of many mesons of different type up to a contribution of the $10$-meson, see Fig.~\ref{fig:bound_state}(b).
After the collision a significant weight of the initial exotic meson is transformed into $1$-mesons, as one might already expect from the spectra in Fig.~\ref{fig:bound_state}(c).

\begin{figure}[tb] 
	\centering
	\includegraphics[width=1.0\linewidth]{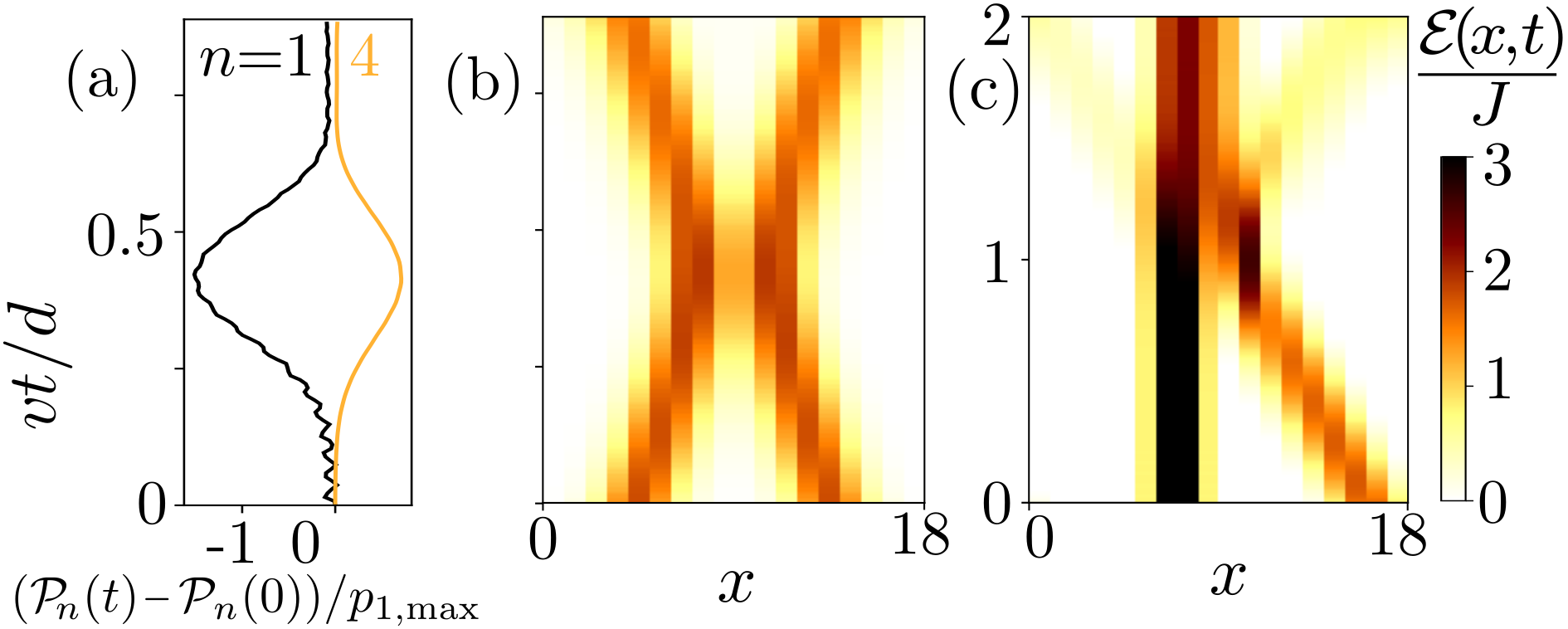}
	\caption{
	Elastic collisions in the quantum Ising chain.
		(a,b) Resonant collision for two $1$-mesons at $h_z=J$.
		(c) Non-resonant collision between one $2$- and one $1$-meson at $h_z=1.2J$.
		(a) Time-dependence of the normalized $n$-meson production $(\mathcal{P}_n(t)-\mathcal{P}_n(0))/p_{1,\mathrm{max}}$.
		(b,c) Spatiotemporal evolution of the local energy $\mathcal{E}(x,t)/J$.
		For the non-resonant collision (c) particle contents don't change over time.
		For both collisions $L=18$, $h_x=0.1J$; the 1-mesons have initial momenta $k=\pi/2$ and the $2$-meson starts at rest.
	}
	\label{fig:elastic}
\end{figure}

\noindent \emph{Elastic collisions.--}
The considered Ising spin chain provides us with the tunability to not only consider inelastic collisions, but also elastic ones, as we show in Fig.~\ref{fig:elastic} for two representative examples.
In Fig.~\ref{fig:elastic}(a,b) we show the collision of two $1$-mesons, again for $h_z=J$.
This collision appears clearly elastic with the outgoing particles identical to the incoming ones.
It is, however, the full spatiotemporal access we have to the dynamics, which allows us to resolve details of the collision event and moreover to obtain information about the mesonic interactions even in case we would not have had prior knowledge on the microscopic Hamiltonian.
Specifically, as soon as the two particles get close to each other reaching a separation $d=2$, the two $1$-mesons fuse to create a $4$-meson, as described by the process induced by the first line of Eq.~(\ref{eq.Hint}).
This fused $4$-meson appears as an intermediate state evolving in a later stage back into two $1$-mesons as the outgoing particles.
This spatiotemporal resolution of the individual meson occupations tells us that the interaction between $1$-mesons is mediated by a $4$-meson as the exchange particle.
As opposed for instance to the case of photon-mediated Coulomb interactions the present exchange particle is massive with a rest mass $m_4$ coupled resonantly to the two $1$-mesons.
Another class of elastic collisions, which can be realised in the proposed setup, is representatively shown in Fig.~\ref{fig:elastic} where we collide a $1$- and a $2$-meson at $h_z/J=1.2$.
In this case the incident projectile splits into a transmitted and reflected contribution upon shifting the heavier $2$-meson due to the Newton cradle effect we have observed already in Fig.~\ref{fig:3+1_spatiotemporal}.
Monitoring the individual meson occupations we see that this class of collisions does not involve a resonant channel and the number of the initial $1$- as well as $2$-mesons remains constant throughout the collision.

\noindent \emph{Concluding discussion.--}
In this work we have shown that particle collisions can be studied in paradigmatic quantum spin chains providing access not only to the asymptotic particle production but also to the full spatiotemporal dynamics of the collision event.
A key advantage of our proposed setup is that it allows for a natural experimental implementation in quantum simulator platforms.
In this context systems of Rydberg atoms appear especially suitable, see also Ref.~\cite{supplement} for more details.
Since they naturally implement the desired quantum Ising chain~\cite{Bernien:2017,Zeiher:2017,Marcuzzi:2017,Lienhard:2018,Guardado:2018,Leseleuc:2018}, the central question is how the achievable coherence times compare with time scales needed to see the two particles colliding.
Fortunately, the collision time scale $t^*=d/u$ can tuned both via relative velocity $u$ of the two participating mesons and the initial distance $d$.
For the $1+1$ collision, say, this gives $u = 2v$ with $v=2 h_x^2/(3J)$ the velocity of a single $1$-meson and in the experimentally relevant units $Jt^*/h=Jt^*/(2\pi\hbar)=(3d/8\pi) (J/h_x)^2 \approx 9.2$ for $d=7$ and $h_x/J=1/3$, which appears within reach of current technology~\cite{Bernien:2017,Zeiher:2017}.
Here, $h$ and $\hbar$ denote Planck's constants and we show a simulation in this parameter regime in Ref.~\cite{supplement}.
Systems of Rydberg atoms provide also the local control to initialize directly the static $n$-mesons by flipping $n$ spins in an otherwise ferromagnetic background~\cite{Marcuzzi:2017,Bernien:2017}.
Generating propagating particles require an alternative preparation as compared to Eq.~(\ref{eq:psin}).
Instead, one can just impose a single spin flip creating a $1$-meson, which will then experience a simple quantum walk, see Eq.~(\ref{eq.H0}), leading to two ballistically propagating fronts with velocity $v$~\cite{supplement}.
Further, the measurement outcomes in systems of Rydberg atoms are spin configurations~\cite{Bernien:2017,Zeiher:2017,Marcuzzi:2017,Lienhard:2018,Guardado:2018,Leseleuc:2018}, from whose statistics one can also directly obtain the individual meson occupations.
While the motivation for the phenomena we have explored originates from high-energy physics, our model is not directly connected to a relativistic quantum field theory by taking a continuum limit.
Importantly, however, there exists an exact mapping to a $\mathbb{Z}_2$ lattice gauge theory~\cite{Lerose:2019jrs}.
Consequently, our results appear relevant for particle collisions as relevant for \emph{lattice} gauge theories in a large mass limit with the key advantage of gaining access to their full spatiotemporal dynamics, which is challenging otherwise.

As we have argued in the context of Fig.~\ref{fig:elastic}, the spatiotemporal information about the collision event can also be used to obtain insights into the interaction channels of the mesons.
For instance, from the collision of the two $1$-mesons in Figs.~\ref{fig:elastic}(a,b) one can directly read off that the effective interaction between these composite particles is mediated by a $4$-meson as the exchange particle.
From this perspective particle collisions as studied here might be also useful in a more general context, lattice gauge theories in particular, in order to gain direct information about effective interactions of the involved composite particles.

\emph{Note added.--} 
During the completion of this project, we became aware of a related work by F.~M.~Surace and A. Lerose, which will appear on the arXiv on the same day.

\textit{Acknowledgements.} Valuable discussions with Johannes Zeiher are gratefully acknowledged. Exact diagonalization was performed using the ELPA library \cite{ELPA:2011,ELPA:2014}. We thank for the help by the support team of MPCDF in Garching and, in particular, Andreas Marek. P.~K. acknowledges the support of the Alexander von Humboldt Foundation. The Gravity, Quantum Fields and Information group at the Max Planck Institute for Gravitational Physics (Albert Einstein Institute) is supported by the Alexander von Humboldt Foundation and the Federal Ministry for Education and Research through the Sofja Kovalevskaja Award.
This project has received funding from the European Research Council (ERC) under the European Union’s
Horizon 2020 research and innovation programme (grant agreement No. 853443), and M. H. further acknowledges support by the Deutsche Forschungsgemeinschaft via the Gottfried Wilhelm Leibniz Prize program.


\bibliography{literature}


\clearpage

\appendix
\section{SUPPLEMENTARY MATERIALS}
\renewcommand{\theequation}{S\arabic{equation}}
\setcounter{equation}{0}

\renewcommand{\thefigure}{S\arabic{figure}}
\setcounter{figure}{0}

\noindent \emph{Effective perturbation theories.--} For weak transverse fields it is natural to decompose the Ising Hamiltonian as follows:
\begin{equation}
\begin{aligned}
&H=H_0+V, \\
&H_0=-\sum _j\left(J \sigma _j^z\sigma _{j+1}^z+h_z\sigma _j^z\right),\\ &V=-h_x\sum _j\sigma _j^x.
\end{aligned}
\end{equation}
For such a decomposition, one can find the Schrieffer-Wolff (SW) generator $S$ that rotates the Hamiltonian $e^{[S,]}H$ to a basis where the off-diagonal terms not preserving $H_0$ are eliminated order by order in powers of $V$~\cite{Motrunich:2017}. At the leading order, one needs to satisfy the condition $[S_1,H_0]+V=0$. The solution, $S_{i,j} = V_{i,j}/(E_i-E_j)$, written in spin basis is
\small
\begin{equation}
\begin{aligned}
    &S_1 = \frac{h_x}{2h_z+4J} P_{j-1}^\uparrow \sigma_j^+ P_{j+1}^\uparrow
    + \frac{h_x}{2h_z-4J} P_{j-1}^\downarrow \sigma_j^+ P_{j+1}^\downarrow\\
    &+ \frac{h_x}{2h_z} P_{j-1}^\downarrow \sigma_j^+ P_{j+1}^\uparrow
    + \frac{h_x}{2h_z} P_{j-1}^\uparrow \sigma_j^+ P_{j+1}^\downarrow -h.c.
    \end{aligned}
\end{equation}
\normalsize
and gives rise to the second-order Hamiltonian $H_2=[S_1,V]/2$ of the following form:
\begin{equation}
\begin{aligned}
H_2&=\sum _j
-\Delta_+P_{j-1}^{\uparrow }\sigma _j^zP_{j+1}^{\uparrow }
-\Delta_-P_{j-1}^{\downarrow }\sigma _j^zP_{j+1}^{\downarrow
}\\
&-\Delta_0\left(P_{j-1}^{\uparrow }\sigma _j^zP_{j+1}^{\downarrow }+P_{j-1}^{\downarrow }\sigma _j^zP_{j+1}^{\uparrow }\right)\\
&+\left(\Delta_+-\Delta_0\right)P_{j-1}^{\uparrow }\left(\sigma _j^+\sigma _{j+1}^-+h.c.\right) P_{j+2}^{\uparrow }\\
&+\left(\Delta_0-\Delta_-\right)P_{j-1}^{\downarrow
}\left(\sigma _j^+\sigma _{j+1}^-+h.c.\right) P_{j+2}^{\downarrow }\\
&+\left(
\Delta_--\Delta_0\right)
P_{j-1}^{\downarrow }\left(\sigma _j^+\sigma _{j+1}^+ + h.c.\right)P_{j+2}^\downarrow.
\end{aligned}
\end{equation}
In the above equation, $\Delta_0\equiv h_x^2/2h_z$ is the string fluctuation energy and $\Delta_\pm\equiv h_x^2/(2h_z\pm 4J)$ is the (false-)meson fluctuation energy in (false-)vacuum. 
Accordingly the wave-function should also be rotated as $\ket{\psi}\to e^S\ket{\psi}=\ket{\psi} + S_1\ket{\psi}+\cdots$. 

Since the domain wall oscillation has been removed by the SW rotation, the classical Ising configurations are conserved in the effective Hamiltonian, which can be straightforwardly measured by the diagonal Ising projectors. When translated back to the original Ising basis, one has to perform an inverse SW rotation to dress the classical configurations. 
The meson energies of the generic $n$-meson up to second order correction are
\begin{equation}
\begin{aligned}
    m_1 =& 4J+2h_z + 2(\Delta_+ - \Delta_0)+ O(h_x^4/J^3),\\
    m_{n\geq 2} =& 4J+2n h_z + (n+2)\Delta_+ ,\\ 
    &+ (n-2)\Delta_- + O(h_x^4/J^3),
    \end{aligned}
\end{equation}
where we followed the same conventions as in the main text. Projecting onto the meson basis, we get the effective Hamiltonian of the following form:
\begin{equation}
\begin{aligned}
&H_{\text{eff}} = 
\sum_{n,x}
m_n\psi_x^{[n]\dagger}\psi_x^{[n]}\\
&-\left(\Delta_0-\Delta_+\right)
\sum_x \left(\psi_x^{[1]\dagger} \psi_{x+1}^{[1]}+h.c.\right)\\
&+(\Delta_- - \Delta_0)
\sum_{m,n,x} \left(\psi_x^{[m+n+2]\dagger} \psi_{x}^{[m]}\psi_{x+m+2}^{[n]}+h.c.\right)\\
&+\left(\Delta_0-\Delta_-\right)
\sum_{m,n,x} 
\left(\psi_{x}^{[m-1]\dagger}\psi_{x+m}^{[n+1]\dagger}
\psi_{x}^{[m]}\psi_{x+m+1}^{[n]}+h.c.\right)\\
&+
(2\Delta_0-\Delta_-)\sum_{m,n,x} 
\left(\psi_{x}^{[m]\dagger}\psi_{x}^{[m]}
\psi_{x+m+1}^{[n]\dagger}\psi_{x+m+1}^{[n]}+h.c.\right).
\end{aligned}
\end{equation}
Physically, the first term describes the rest mass of mesons; the second term describes the hopping of 1-meson; the third term describes the fusion of two strings as well as the reverse process of the breaking of a string; the fourth term describes a repulsive nearest-neighbour density interaction. 
Notice that $h_z=2J$ is a singular case with a first-order resonance in $h_x$, which we exclude in our consideration.  
The last term does not commute with $H_0$ and and can be further eliminated by a higher-order Schrieffer-Wolff generator, unless $h_z\simeq J$. This is the term that was not considered in Ref.~\cite{Motrunich:2017}, as the authors in the work have been focusing on the limit of weak longitudinal fields. 
However, in the following we aim to concentrate on $|h_z-J|\sim O(h_x^2/J)$, where the coupling constants to second order are reduced to $\Delta_0\approx 1/2$, $\Delta_+\approx 1/6$, $\Delta_-\approx -1/2$ in the unit of $h_x^2/J$. Correspondingly, $m_1 \approx 4J + 2h_z -h_x^2/3J $, $m_{n\geq2}\approx 4J+2n h_z + (4-n)h_x^2/3J $, which reproduces the effective Hamiltonian results quoted in the main text.

\begin{figure}[h] 
   \centering
   \includegraphics[width=\columnwidth]{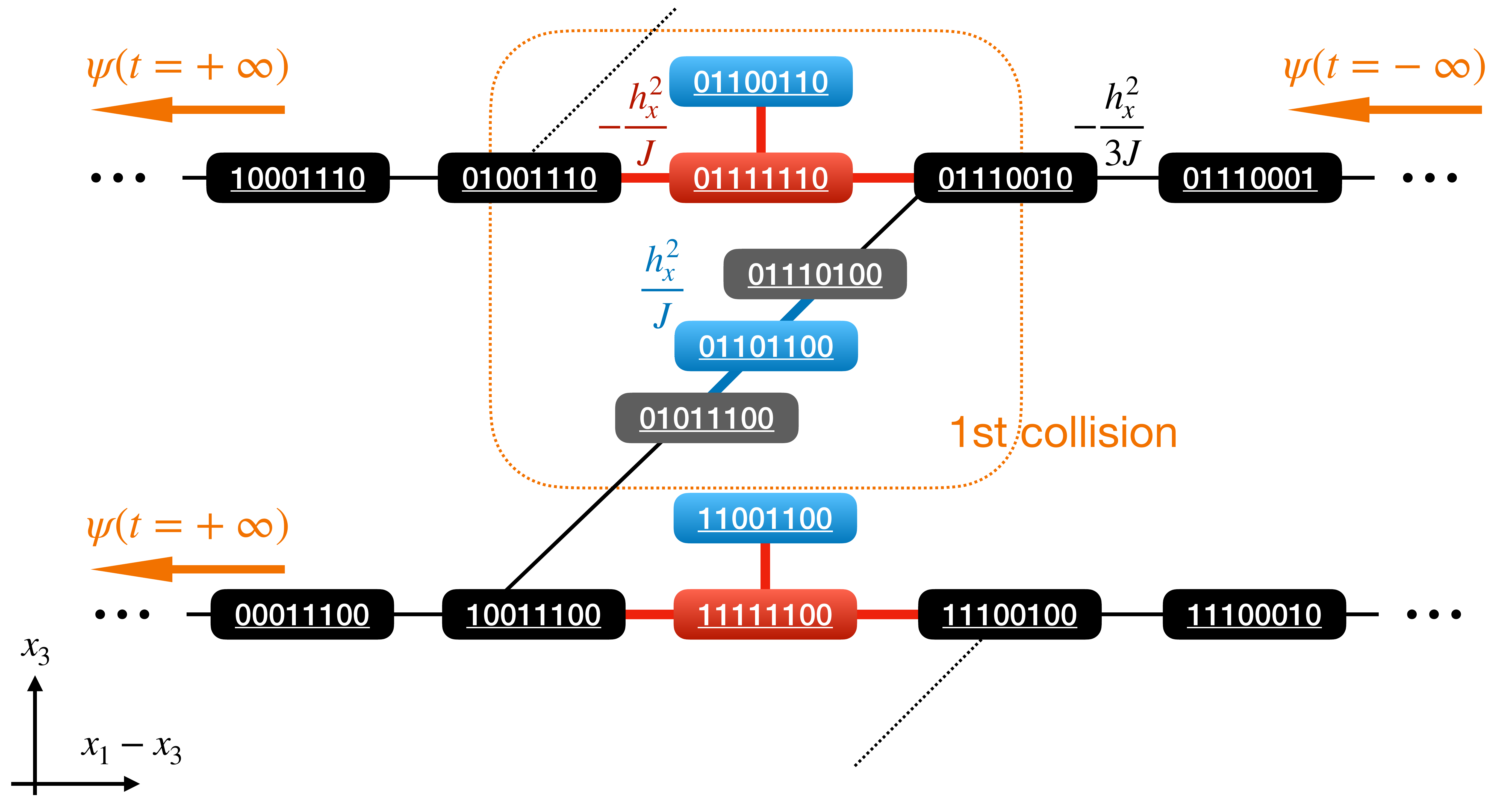} 
   \caption{A comb shaped graph illustrates the Ising levels relevant in the 3+1-meson collision processes, and their connectivity under the 2$^{\mathrm{nd}}$ order effective Hamiltonian. The levels are organized by the relative position between 3-meson and 1-meson $x_1-x_3$ and the position of 3-meson $x_3$ as noted in the figure, where 0(1) stands for spin up(down) respectively, and the hopping strengths are denoted along side with the bond. Black blocks are for isolated 3-meson and 1-meson; gray blocks denote the nearest neighbour 3-meson and 1-meson pair; blue blocks are for 2+2-meson bound state; red blocks are for 6-meson. Orange arrows guide the dynamical process of 3+1-meson collision, which in general lead to multiple outgoing channels. }
   \label{fig:levelGraph}
\end{figure}

While the effective Hamiltonian above might look rather complex in comparison with the original Ising model, for the collision protocol described in the main text it involves only a subspace with polynomial complexity.

The graph of states relevant for the $3+1$-meson is shown in Fig.~\ref{fig:levelGraph}. The dynamical process of $3+1$-meson collision is equivalent to the propagation through this comb-like effective graph of a single quantum particle, which is initially located in the far right hand side of the first row. As denoted by the orange arrow, it moves to the left and has two channels: (i) tunneling across the red block, which means that the 3- and 1-mesons are fused into a 6-meson and its resonant next nearest neighbour 2-meson-pair bound state; (ii) tunneling across the blue block, which means that the 3-meson and 1-meson exchange their electric string lengths leading to a nearest neighbour 2-meson-pair bound state. 
After the first collision, the 1-meson tunnels to the left, and then faces the choice of either propagating away to the left, or returning to combine with the 3-meson again. It is the latter choice that leads to the effective diffusion of 6-meson or 2-meson-pair bound states, which could not move by themselves under second order perturbation. 
It should be emphasized that the blue and red blocks have an energy barrier comparable with the bond strength, all on the energy scale of $O(h_x^2/J)$. Taking the black blocks for isolated 3+1-meson configurations as a reference, the energy barrier i.e. the relative rest mass of the $111111$ block is $\epsilon_b=m_6-m_3-m_1=4 (h_z-J) - 2h_x^2/(3J)$, and that for $11011$ block is $\epsilon_d=3h_x^2/2J+2m_2-m_3-m_1=17h_x^2/6J$, with the repulsive density interaction taken into account. The bond strength is outlined in Fig.~\ref{fig:levelGraph}. Therefore, the effective kinetic energies of the 6-meson and 2-meson-pair bound states, mediated by the 1-meson, are not exponentially suppressed. This could be understood because the 1-meson, while being massive with respect to the Ising vacuum, is relatively massless with respect to the relevant energy shell. 

As this and other graphs encapsulating our collision protocols involve only a polynomial number of states, our numerical calculations were able to deal with large system sizes considered in the main text.

The 7+1-meson collision involves a much more complex graph with the backbone of 4-dimensional square grid since it contains four light mesons.

\noindent \emph{Exotic meson bound states.--} Due to the conspiracy of strong confinement and weak quantum fluctuation, there exist multiple exotic heavy meson bound states lying outside of the light meson continua. By exact diagonalization we obtain the spectrum and the eigenstates in the zero-momentum sector. While the spectrum has been shown in the main-text, here we give the eigenstates of these exotic bound states in the mesonic basis. 

For the 6-meson shell relevant to 1+3-collisions, we have three bound states (indicated by green lines in Fig.~2d) resonant with 6-meson: 
\begin{equation}
\begin{aligned}
    \ket{6^+_b} &\approx 0.78\ket{111111} +0.57\frac{\ket{111001}+\ket{100111}}{\sqrt{2}}\\
    &+ 0.23\ket{110011} + \ldots,\\
    \ket{6^+_c} &\approx 0.24\ket{111111} - 0.63\frac{\ket{111001}+\ket{100111}}{\sqrt{2}}\\
    &+ 0.52\ket{110011} + \ldots,\\
    \ket{6^+_t} &\approx 0.49\ket{111111} - 0.37\frac{\ket{111001}+\ket{100111}}{\sqrt{2}}\\
    &- 0.78\ket{110011} +\ldots,
\end{aligned}
\end{equation}
where the ellipsis contains in particular weak contributions from 1+3 configurations at larger separations and $\ket{11011}$ configuration. The states energies are respectively $(E-m_6)/E_{1,\mathrm{kin}}=-2.015, 2.304, 3.954$, where the lower index d,~c, and~t stand for the bottom, center, and top levels as seen in Fig.~2(d) in the main text.

Therefore, it is not only the 6-meson, but also the inversion-symmetric 3+1-meson pair and the 2+2-meson pair have significant overlap with these bound states. They all affect the dynamics of the collision in a non-perturbative way, as the coupling between the 6-meson (2+2-meson pair) and the 3+1-meson pair is on the same scale as the velocity of 1-meson and, moreover, the 6-meson (2+2-meson) bound state can diffuse via the propagation of the 1-meson.

For the 7-meson shell, there are two bound states significantly overlapping with the 7-meson,
\begin{align}
&\ket{7^+_b} \approx 0.44\ket{1111111}
+ 0.46\ket{1001001} \nonumber\\
&+0.49\frac{\ket{1111001}+\ket{1001111}}{\sqrt{2}} \nonumber\\ &+0.15\frac{\ket{1110011}+\ket{1100111}}{\sqrt{2}}+ \ldots,
\label{eq:7-meson}
\end{align}
\begin{align}
&\ket{7^+_t} \approx 0.49\ket{1111111}
+ 0.25\ket{1001001} \nonumber\\
&-0.46\frac{\ket{1111001}+\ket{1001111}}{\sqrt{2}}\nonumber\\
&-0.62\frac{\ket{1110011}+\ket{1100111}}{\sqrt{2}}+ \ldots,
\label{eq:7-meson_t}
\end{align}
with energies $(E-m_7)/E_{1,\mathrm{kin}}=-3.135$ and $-1.371$ respectively. The bottom meson level is separated from the (1+1+1)-meson continuum by a small gap $\Delta E/E_{1,\mathrm{kin}}=-0.135$. The remaining spectral weight of 7-mesons hybridises with the (1+1+1)-meson continuum band, see Fig.~3(c) in the main text.

\begin{figure}[tbh] 
	\centering
	\includegraphics{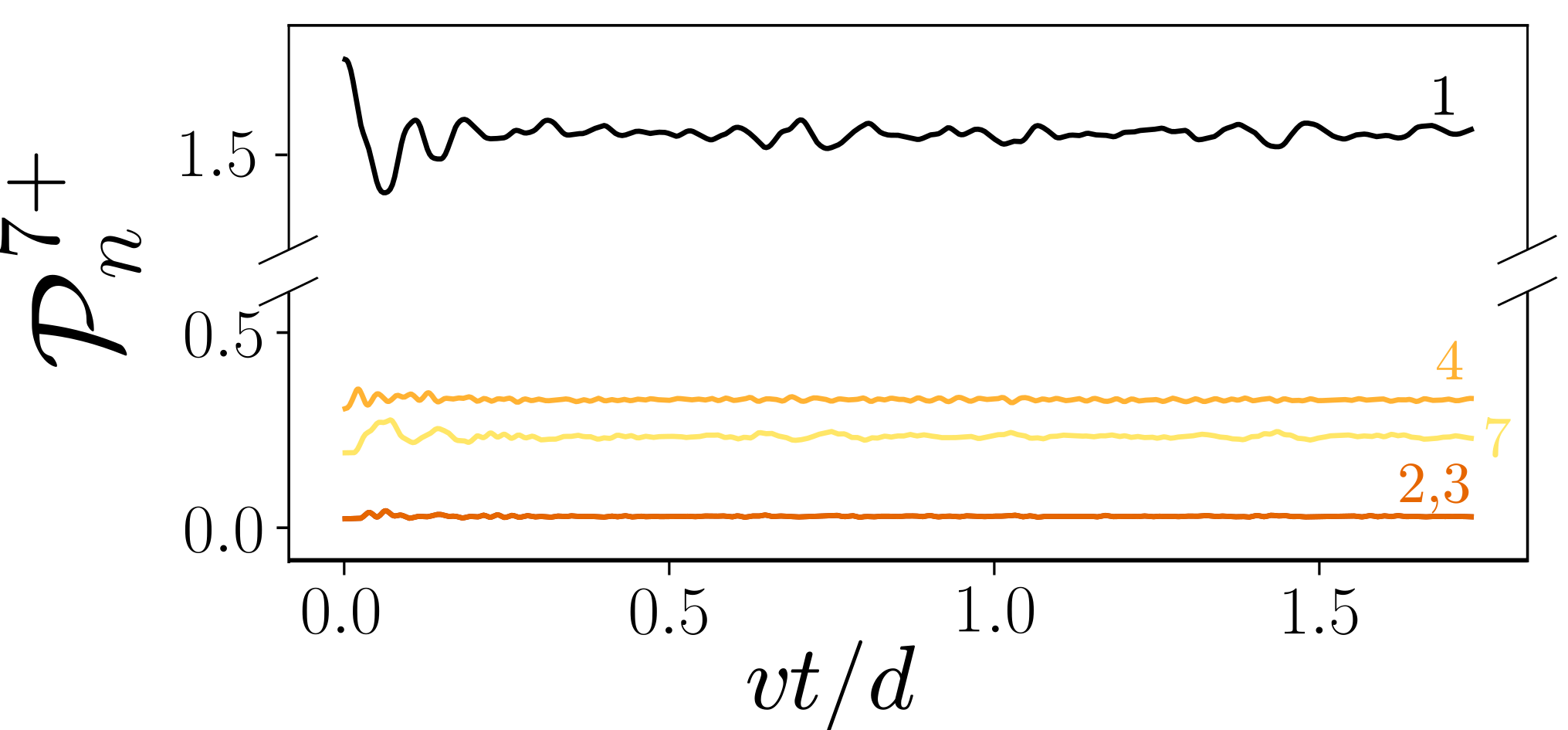}
	\caption{
	Stability of the approximately prepared exotic $7^+$-meson bound state $\ket{7^+_b}$, eq. (\ref{eq:7-meson}): time dependence of projectors $\mathcal{P}^{7+}_n$ for $n=1$-$4,7$. Here $d=L/2=50$, similarly to Fig.~3 in the main text.
	}
	\label{fig:7-meson_stability}
\end{figure}

Since the $7^+$-mesons are significantly heavier than single 1-mesons, the zero-momentum sector amplitudes (\ref{eq:7-meson},\ref{eq:7-meson_t}) give also a good approximation to the localized states. In Fig.~\ref{fig:7-meson_stability} we explicitly check for the stability of the approximately found $\ket{7^+_b}$ meson used in the main text.

\noindent \emph{Other choices of resonant longitudinal fields.--} Beyond $h_z\simeq J$, a sequence of weaker longitudinal fields could also satisfy the classical resonance condition at 0-th order of $h_x$~\cite{Verdel:2020}. This can be achieved when the total rest mass of a pair of domain walls is commensurate with the string tension i.e. $h_z\simeq 2J/l$ for $l=1,2,3,\ldots$, where $l$ is the length of the broken string. For weaker longitudinal field, the expense is that higher-order perturbations are required to transform a pair of domain walls into an electric string, which in general is suppressed at short times. 

\begin{figure}[tbh] 
	\centering
	\includegraphics{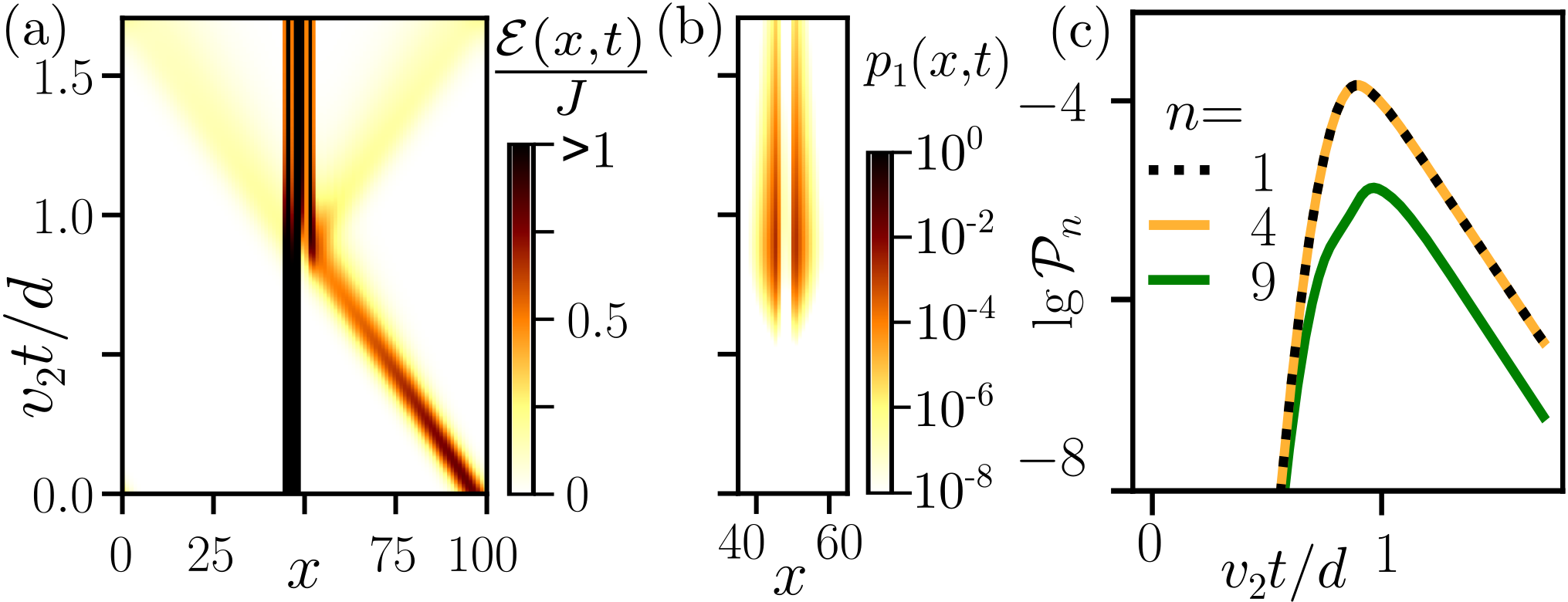}
	\caption{
	    4-th order nearly resonant 3-meson+2-meson collision for $h_z/J=1/2$. 
	    Spatiotemporal evolution of: (a) local energy $\mathcal{E}(x,t)$; (b) local 1-meson projector $p_1(x,t)$.
	    (c) Logarithm of the integrated projectors $\mathcal{P}_n(t)$ for $n=1,4,9$.
	    The data is obtained by $4^{\mathrm{th}}$-order perturbation theory; $v_2=8h_x^4/3J^3$ is the velocity of 2-meson at $k=\pi/2$. The classical value $h_z/J=1/2$ is slightly off-resonant due to $2^{\mathrm{nd}}$-order corrections.
	}
	\label{fig:hz=0.5_resonance}
\end{figure}

In the following we consider a promising candidate of a resonant value $h_z=0.5J$ corresponding to flipping $l=4$ spins, so the resonant conversion between domain wall pairs and electric strings takes place on the order $O(h_x^4/J^3)$, which concurs with the velocity of 2-mesons. Therefore we use the 2-meson as the light incident particle, and the 3-meson as a static target, and we investigate the dynamical process on the time scale $O(J^3/h_x^4)$. 
The resonant channel 3+2 $\rightarrow$ 9-meson $\rightarrow$ 4+1 allows metastable 1-, 4-, 9-mesons with the finite lifetimes, see Fig.~\ref{fig:hz=0.5_resonance}. Qualitatively similar dynamics is observed for this 3+2-meson collision in $h_z\simeq 0.5 J$ as the $h_z\simeq J$ 3+1-meson collision in the main text.

\noindent \emph{Collision protocol relevant for Rydberg atom experiment.--} While the limit of asymptotically weak transverse fields allows us to gain analytical insights, the propagation of the mesons also becomes extremely slow leading to a long waiting time for the collision event to happen. For potential experiments this would be, of course, disadvantageous, so that in the following we will explore the dynamics at stronger transverse fields and closer initial distance between target and incident particles. Also, instead of creating the wavepackets which require superposition states we initialize the particles by flipping the spins on top of the classical ground state. While higher-meson bound states are still static objects on the considered time scales for the stronger fields, the $1$-meson, created by flipping a single spin, splits mainly into two peaks propagation in opposite directions with a velocity $v$ at momentum $k=\pi/2$, consistent with the expected quantum walk in the limit $h_x/J\ll1$, see Eq.~(\ref{eq.H0}) in the main text.

\begin{figure}[tbh] 
	\centering
	\includegraphics{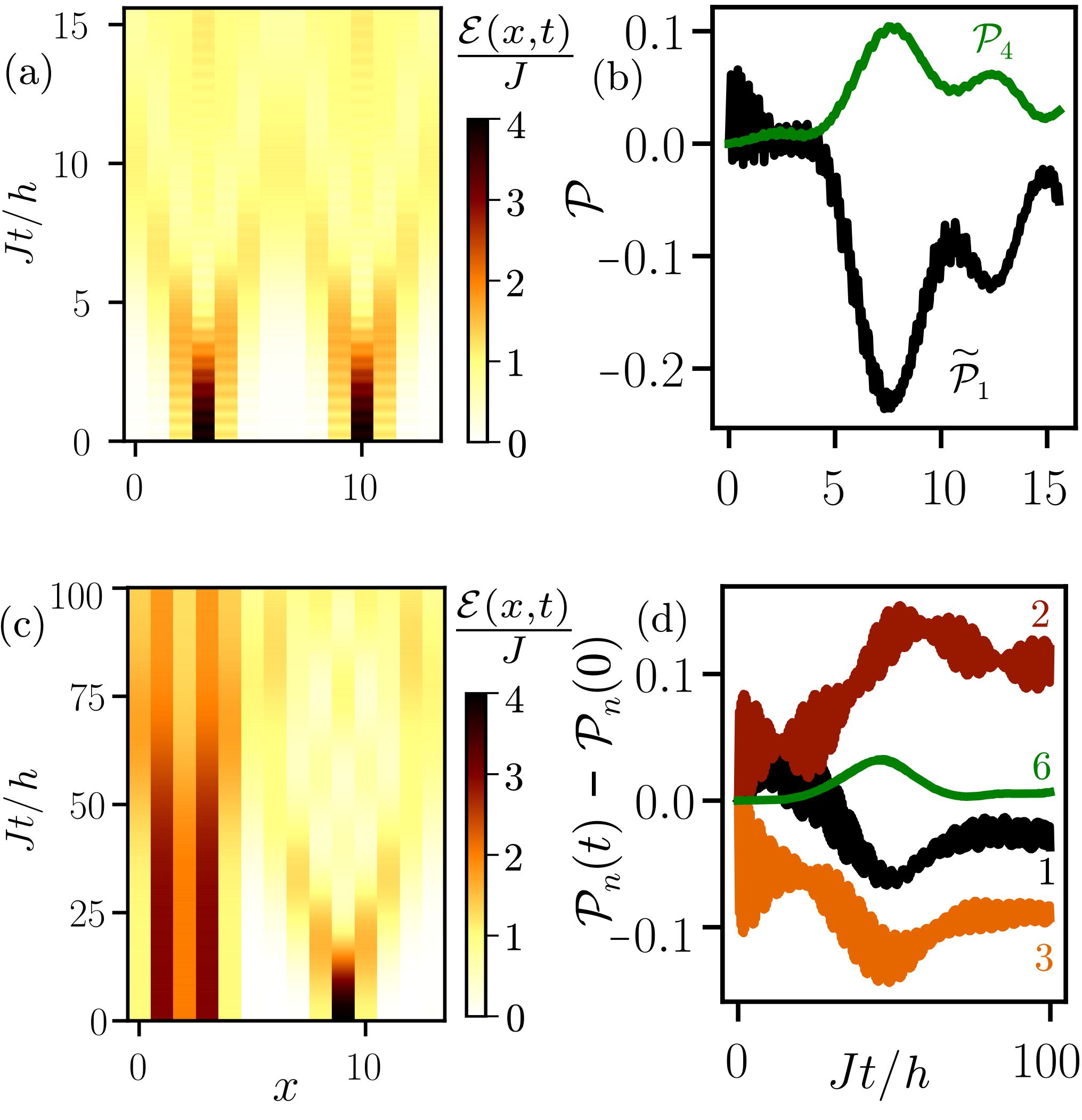}
	\caption{
	    Particle collisions at stronger transverse fields as experimentally relevant with all mesons created by spin flips instead of wavepackets as studied in the main text. 
	    (a,b) 1+1-meson collision for $h_x/J=1/3$.
	    (c,d) 3+1-meson collision for $h_x/J=0.15$.
	    (a,c) Spatiotemporal evolution of local energy $\mathcal{E}(x,t)$. (b,d) Time-dependence of the particle production: 
	    (b) dressed 1-meson projector $\tilde{P}_1(t)=P_1(t)+P_2(t)-P_1(0)-P_2(0)$ and 4-meson projector $P_4(t)$;
	    (d) $\mathcal{P}_n (t)-\mathcal{P}_n(0)$ for $n=1,2,3,6$. 
	    The data is generated using the full Ising Hamiltonian; initial separation between the centers of the mesons is $d=7$, $h_z=J$, $L=14$; $h$ denotes the Planck's constant.
	}
	\label{fig:experimental}
\end{figure}

In Fig.~\ref{fig:experimental}(a,b) we consider the 1+1-meson collision similar to Fig.~\ref{fig:elastic}(a) of the main text obtained using exact diagonalization of the full Ising Hamiltonian for $h_x/J=1/3$, $h_z=J$, $L=14$, and initial separation $d=7$ between the 1-mesons. We give time in experimental units relative to Planck's constant $h$. The single spin-flips in the initial condition lead to propagating wave packets of $1$-mesons impacting onto each other resulting in a collision occurring on a time scale $Jt^*/h\approx 10$ which is within the experimentally accessible regime~\cite{Zeiher:2017}. The collision leads to changes in the meson occupations, as depicted in Fig.~\ref{fig:experimental}(b). As one can see, the collision leads to the production of $4$-mesons at the expense of a reduction of $1$-mesons. Note, that we measure the $1$-meson occupation in a slightly different way as compared to the main text where we consider weaker transverse fields. As already discussed in the main text, the measurement of the $n$-mesons by the projectors $\mathcal{P}_n(t)$ is only exact in the asymptotic limit $h_x/J\to0$ with corrections at nonzero transverse fields. For the case considered in Fig.~\ref{fig:experimental}(a,b) we find from a perturbative analysis taking into account $h_x/J$ corrections that the $1$-mesons are dressed by two mesons. As a consequence, we can get a better estimate $\tilde{\mathcal{P}}_1(t)=\mathcal{P}_1(t)+\mathcal{P}_2(t)$ of the actual $1$-meson occupation by adding up the two bare $1$- and $2$-meson contributions.

In Fig.~\ref{fig:experimental}(c,d) we explore the more complex $3+1$-meson collision analogous to the one studied in Figs.~\ref{fig:3+1_spatiotemporal},\ref{fig:3+1_projectors}. Since the collision involves more types of mesons, namely 1-, 2-, 3-, and 6-meson, it becomes more important to use a smaller value of $h_x/J=0.15$ in order to be able to neglect the dressing contributions to the mesons. Due to smaller $h_x$, the characteristic collision time is now larger, $Jt/h\sim 50$, which is larger by a factor of $\sim 4$ as compared to the timescales reached in the recent experiment~\cite{Zeiher:2017}. This appears challenging, but might be achievable in the foreseeable future.

\end{document}